\documentclass{article}[12pt]

\usepackage{epsfig}
\usepackage{amsmath}
\usepackage{hyperref}
\usepackage{mathbbol}
\usepackage{mathrsfs}
\usepackage{subfig}
\usepackage{lscape}
\usepackage{vmargin}
\setpapersize{A4}
\setmargrb{25mm}{25mm}{25mm}{25mm}

\newcommand{\be}{\begin{equation}}
\newcommand{\ee}{\end{equation}}
\newcommand{\ba}{\begin{equation} \begin{aligned}}
\newcommand{\ea}{\end{aligned} \end{equation}}
\newcommand{\baa}{\begin{equation} \begin{align}}
\newcommand{\eaa}{\end{align} \end{equation}}

\newcommand{\elsum}[1]{\left|\left|#1\right|\right|}

\newcommand{\Reals}{\mathbb{R}}
\newcommand{\Integers}{\mathbb{Z}}
\newcommand{\unitop}{\mathbb{1}}

\newcommand{\SIR}{\textit{SIR}}

\newcommand{\bra}[1]{\left< #1 \right|}
\newcommand{\ket}[1]{\left| #1 \right>}
\newcommand{\ip}[2]{\left< #1 | #2 \right>}

\title{\bf Exact epidemic dynamics for generally clustered, complex networks}
\author{Thomas House\\
Mathematics Institute, University of Warwick, Coventry, CV4 7AL.}
\date{}

\begin{document}

\maketitle

\begin{abstract}

The last few years have seen remarkably fast progress in the understanding of
statistics and epidemic dynamics of various clustered networks. This paper
considers a class of networks based around a concept (the \textit{locale}) that
allows asymptotically exact results to be derived for epidemic dynamics.  While
there is no restriction on the motifs that can be found in such graphs, each
node must be uniquely assigned to a generally clustered subgraph to obtain
analytic traction.

\end{abstract}

\section{Introduction}

Recent progress on analytic approaches to epidemics on clustered networks
has been extremely fast.  Models have been proposed based on
households~\cite{Trapman:2007,Ball:2009, Ball:2010}, and the more general
concept of local-global networks~\cite{Ball:2002,Ball:2008}.  Another recent
innovation has come from generalisations of random graph
theory~\cite{Newman:2009,Miller:2009,Gleeson:2009,%
Gleeson:2009a,Karrer:2010,Bollobas:2011},
and at the same time, general methods have been proposed for manipulation of
master equations~\cite{Simon:2011,Ross:2010}. These complement the traditional
epidemiological approach to clustering based on moment
closure~\cite{Keeling:1999} that has recently been applied to graphs with more
general motif structure~\cite{House:2009}.

This paper draws on much of this recent activity, making three main
contributions. Firstly, a set of networks is defined using the concept of a
locale (which is distinct from the recently introduced concept of a
role~\cite{Karrer:2010} but closely related to the hypergraphs
of~\cite{Bollobas:2011}) that have no restriction on the motifs that can be
present. Secondly, epidemic dynamics are derived for these networks---the first
time that manifestly asymptotically exact results for transient epidemic
dynamics of an large clustered network with non-homogeneous mixing outside the
clusters have been derived. Finally, techniques are presented for practical
efficient calculation of quantities of interest.

\section{General Theory}

\subsection{Network generation}

We start with the definition of a network (or graph---we use the terms
interchangeably) $G$ of size $N$ as a set of nodes (vertices) $V$, and a set of
links (edges) $E \subseteq V \times V$.  Using $\Integers_N$ to stand for the
set of integers from 1 to $N$, nodes are indexed by $i, j, \dots \in
\Integers_{N}$. The information contained in a network can be encoded in an
adjacency matrix $\mathbf{A} = (A_{ij})$, whose elements are given by
\be
A_{ij} = \begin{cases}
1 & \text{ if } \left( i , j \right) \in E \text{ ,}\\
0 & \text{ otherwise.}
\end{cases}
\ee
Here we consider symmetric, non-weighted networks without self-links and so
$A_{ii} = 0$, $A_{ij} = A_{ji}$.

We now present a model for network creation that has been essentially
considered in the context of random graph theory~\cite{Bollobas:2011}. This
starts by defining a set of objects we call here \textit{stubby subnets}, which
are indexed by type $\sigma$. A stubby subnet of type $\sigma$ and size
$n_\sigma$ consists of three elements:
\begin{enumerate}
\item A set of nodes $v^\sigma \cong \Integers_{n_{\sigma}}$;
\item A set of within-subnet links, $e^\sigma \subseteq v^\sigma \times
v^\sigma$, with a within-subnet adjacency matrix $\mathbf{a}^\sigma$ defined
as for $\mathbf{A}$ above;
\item A vector of `stubs' $\mathbf{s}^\sigma$, such that $\forall i\in
v^\sigma, s_i^\sigma \in \Integers_{+}$.
\end{enumerate}
A full network is then constructed in the following way. We choose a finite
number of stubby subnet types, indexed by $\sigma$, and pick a number $N_\sigma
\gg 1$ of each stubby subnet type.  Using tensor sums $\oplus$ to represent the
aggregation of objects without the removal of `duplicates' that would be
implicit in set-theoretic union, the network size, nodes and first part of the
link set are given respectively by
\be
N = \sum_\sigma N_\sigma n_\sigma \text{ ,} \qquad
V = \bigoplus_{\sigma} \bigoplus_{m = 1}^{N_\sigma} v^\sigma \text{ ,} \qquad
E_1 = \bigoplus_{\sigma} \bigoplus_{m = 1}^{N_\sigma} e^\sigma
\text{ .}
\ee
The remainder of links are then provided by constructing a full vector of
`stubs' and connecting these using the standard Configuration
Model~\cite{MolloyReed} where distinct stubs are uniformly paired at random to
form links.
\be
\mathbf{S} = \bigoplus_{\sigma} \bigoplus_{m = 1}^{N_\sigma} 
 \mathbf{s}^\sigma \text{ ,} \quad
E_2 = \mathrm{ConfigurationModel}(V,\mathbf{S}) \text{ ,} \quad
E = E_1 \cup E_2\text{ .} \label{ssnetgen}
\ee 
In the limit where the network size $N$ is large, the duplicate links and
self-edges produced through this construction should be Poisson distributed and
independent of $N$ as in e.g.~\cite[Theorem 3.1.2]{Durrett:2007} however for
explicit generation of finite-size networks, the simple removal of duplicates
implicit in~\eqref{ssnetgen} is commonly used and should not significantly
alter the epidemic dynamics.

Having defined such a network, it is straightforward to calculate degree
distributions and clustering coefficients, since a node $i$ from a stubby
subnet of type $\sigma$ has degree and clustering coefficient
\be
d_i = \sum_j A_{ij} = s_i^\sigma + \sum_j a_{ij}^{\sigma} \text{ , and} \quad 
\phi_i = \frac{\sum_{j,k} A_{ij} A_{ik} A_{jk}}%
{\sum_{j,k} A_{ij} A_{ik} (1-\delta_{jk})}
= \frac{{(\mathbf{a}^\sigma)^3}_{ii}}{d_i(d_i -1)} \text{ ,}
\ee
respectively.  From consideration of the standard configuration
model~\cite{Durrett:2007,MolloyReed}, if all stubby subnets are internally
connected then a giant component emerges provided
\be
\sum_\sigma M_\sigma D^\sigma ( D^\sigma - 2) > 0
\text{ ,}\quad \text{where} \quad D^\sigma := \sum_{i=1}^{n_\sigma} s_i^\sigma
 \text{ ,} \quad
M_\sigma = \frac{N_{\sigma}}{\sum_{\sigma'} N_{\sigma'}} \text{ .}
\ee

\subsection{Invasion and final size}

We now introduce a framework for the determination of whether a network of the
kind considered can support the invasion of a species obeying
susceptible-infectious-recovered (\SIR{}) dynamics.  To do this, we define the
concept of a \textit{locale}, which is a stubby subnet of type $\sigma$,
together with an `origin' node $o \in v^\sigma$.  Clearly, there are at least
$\sum_\sigma n_\sigma$ such locales to consider, although symmetries may reduce
the effective number of these. Locale types are denoted using indices like
$\lambda = \left( \sigma, o \right)$.

Invasibility of a network of the type under consideration (i.e.\ one
constructed from stubby subnets) can therefore be considered by constructing a
branching process on locales. If we define a `locale next generation' matrix as
the number of secondary locales infected by an initially infected locale early
in the epidemic, then we can use the dominant eigenvalue of such a matrix to
define a threshold parameter.

In order to do this, we need to define two dynamical quantities. The first of
these is $T$, the probability that infection eventually passes across a network
link where one node starts infectious and the other susceptible. The second is
$P_\sigma(j | o)$, which is the probability that within the locale $\left(
\sigma, o \right)$, where infection is first introduced to node $o$, that
infection eventually reaches node $j\in v^\sigma$. The calculation of these two
quantities depends on the precise dynamical system underneath the transmission
process, but once they have been determined, the locale next generation matrix
(interpreted as the expected number of locales of type $\bar{\lambda} = \left(
\bar{\sigma}, \bar{o} \right)$ created by a locale of type $\lambda= \left(
\sigma, o \right)$ early in the epidemic) is given by
\be \mathcal{K}^{L}_{\bar{\lambda} \lambda} = 
T \frac{M_{\bar{\sigma}} s^{\bar{\sigma}}_{\bar{o}}}{%
 s_{\mathrm{tot}}}
 \left( \left(s_o^\sigma -1\right)
  + \sum_{j \in v_\sigma \ominus o} P_\sigma (j | o) s_j^\sigma \right)
 \text{ ,} \label{Kl}
\ee
where the normalised total number of stubs in the network is
\be
s_{\mathrm{tot}} = \sum_{\sigma} M_{\sigma} \sum_{i \in v_{\sigma}}
s_i^{\sigma} \text{ .}
\ee
The expression~\eqref{Kl} is essentially composed of three parts. Before the
bracket is the relative weighting by stubs of susceptible locales of type
$\bar{\lambda}$; within the bracket is firstly the number of stubs in the
origin node of locales of type $\lambda$, subtracting 1 to allow for the
link that passed infection to the origin node; and secondly within the bracket is
the number of stubs attached to other nodes in locales of type $\lambda$
weighted by the probability that those nodes are locally infected by $o$.  The
locale basic reproduction number, which is different from the standard basic
reproductive number $R_0$, is then the dominant eigenvalue of this matrix
\be
 R_L := \elsum{\mathcal{K}^L} \text{ .}
\ee
By using a `susceptibility sets' argument as in~\cite{Ball:2009,Ball:2010}, the
final size of an epidemic can also be calculated using the following set of
transcendental equations:
\ba
R_{\infty} & = 1 - 
 \frac{\sum_{\sigma} M_\sigma \sum_{i\in v_{\sigma}} x_i^\sigma }%
 {\sum_\sigma M_\sigma n_\sigma} \text{ ,}\\
x_i^\sigma & = 1 - P(i|\boldsymbol{\pi}^\sigma) \text{ ,} \\
(\boldsymbol{\pi}^\sigma)_i & = \left( (1-T) + T \sum_{\lambda'} 
 \frac{M_{\sigma'}s_{o'}^{\sigma'}}{s_{\mathrm{tot}}} \tilde{x}_{o'}^{\sigma'}
 \right)^{s_i^\sigma} \text{ ,} \\
\tilde{x}_i^\sigma & = 1 - P(i|\tilde{\boldsymbol{\pi}}^{\sigma,i}) \text{ ,} \\
(\tilde{\boldsymbol{\pi}}^{\sigma,i})_j & = \left( (1-T) + T \sum_{\lambda'} 
 \frac{M_{\sigma'}s_{o'}^{\sigma'}}{s_{\mathrm{tot}}} \tilde{x}_{o'}^{\sigma'}
 \right)^{s_j^\sigma - \delta_{i,j}} \text{ .}
\label{finalsizes}
\ea
Here $R_{\infty}$ is the proportion of the population that is ultimately
infected by the epidemic, while $x_i^\sigma$ is the probability that the $i$-th
node in a stubby subnet $\sigma$ avoids infection during the epidemic.  The
$j$-th element of vector $\boldsymbol{\pi}^\sigma$ is the probability that the
$j$-th node in a stubby subnet $\sigma$ avoids global infection (i.e.\ through
one of its stubs) during the epidemic, with $P(i| \boldsymbol{\pi}^\sigma)$
standing for the probability that a node $i$ in $\sigma$ is ultimately infected
given such a vector.

Each of the equations in~\eqref{finalsizes} can be paraphrased in English: the
first follows essentially from the definitions of the quantities involved, and
states that the final epidemic size is one minus the proportion of nodes that
avoid infection; the second states that nodes avoiding infection are neither
infected locally nor globally; the third states that each global neighbour of a
node that avoids infection either does not transmit along the relevant link, or
avoids infection itself; the fourth states that contacts avoiding infection are
not infected locally, and are not infected globally once the link to the
original node is discounted; and the fifth states that contacts of contacts
avoiding infection either do not transmit along the relevant links, or avoid
infection themselves.

\subsection{Full Dynamics}

In order to consider full transient dynamics for the system, we assume that
transmission of infection across a link is a one-step Poisson process,
happening at rate $\tau$, and that recovery is Markovian with rate $\gamma$.
Our methodology is straightforwardly extended to the case where transmission
happens at a variable rate during an individual's infectious period or the case
of non-exponentially distributed recovery times through the addition of extra
disease compartments; the density of phase-type distributions means that an
arbitrarily good approximation can be made given a sufficiently large state
space~\cite{Neuts:1975}. It is also possible to calculate invasion thresholds
and final sizes for arbitrary recovery times (i.e.\ without approximation by a
phase-type distribution) since these rely only on the probability of
transmission across a global link, $T$ (which is $\tau/(\tau + \gamma)$ in the
Markovian case) and `multitype' final sizes that can also be calculated in
generality~\cite{Ball:1986}.

\subsubsection{Notation}

Since the full model dynamics we consider are rather hard to write down, we
make use of Dirac `bra-c-ket' formalism, using the appropriate links to Markov
chains~\cite{Dodd:2009}, to allow sufficiently compact notation for the full
dynamical system. While this notation may not be familiar to mathematicians
without experience of theoretical physics, it has been used in expository
accounts of e.g.\ vector spaces~\cite{Dennery:1967}.

Let $\mathcal{V}$ be a vector space, with elements $\ket{} \in
\mathcal{V}$ (called \textit{kets}). The inner product (or \textit{bracket}) is
a map
\be
\mathcal{V} \times \mathcal{V} \rightarrow \Reals \text{ ,} \qquad
(\ket{x}, \ket{y}) \mapsto \ip{x}{y} \text{ .}
\ee
A bijection exists between $\mathcal{V}$ and its adjoint vector space
$\mathcal{V}^*$, which has elements $\bra{} \in \mathcal{V}^*$ (called
\textit{bras}). This bijection maps $\ket{x} \in \mathcal{V}$ to $\bra{x}
\in \mathcal{V}^*$, $\forall x$, meaning that the inner product can be used to
define a product
\be
\mathcal{V}^* \times \mathcal{V} \rightarrow \Reals \text{ ,} \qquad
(\bra{x}, \ket{y}) \mapsto \bra{x} \cdot \ket{y} := \ip{x}{y} \text{ .}
\ee
Since the distinction between this product and the inner product does not
matter for performing calculations, all objects of the form $\ip{}{}$ are often
simply called `brackets'.

Operators, written in the form $\hat{\mathcal{O}}$, are maps
\be
\hat{\mathcal{O}}:
\mathcal{V} \rightarrow \mathcal{V} \text{ .}
\ee
Where $\mathcal{V}$ is spanned by a set of basis vectors, such operators can
be defined uniquely through their action on each of this set's elements. A
special class of operators are projections, written in the form $\ket{}\bra{}$,
which are defined to obey
\be
\ket{y}\bra{y}:
\mathcal{V} \rightarrow \mathcal{V} \text{ ,} \qquad
\ket{y}\bra{y} \ket{x} \mapsto \ip{y}{x} \ket{y} \text{ .}
\ee
The final, general notational convention to introduce is tensor multiplication
of vectors and operators.
Where $\ket{x}\in\mathcal{V}$, $\ket{\tilde{x}}\in\tilde{\mathcal{V}}$, and 
$\hat{\mathcal{O}}_1:\mathcal{V}\rightarrow\mathcal{V}$, 
$\hat{\mathcal{O}}_2:\tilde{\mathcal{V}}\rightarrow\tilde{\mathcal{V}}$, 
\be
\ket{x}\otimes\ket{\tilde{x}} \in \mathcal{V} \times
\tilde{\mathcal{V}} \text{ ,}
\qquad 
(\hat{\mathcal{O}}_1\otimes \hat{\mathcal{O}}_2) (\ket{x}\otimes\ket{\tilde{x}})
= (\hat{\mathcal{O}}_1\ket{x})\otimes (\hat{\mathcal{O}}_2\otimes\ket{\tilde{x}})
\text{ .}
\ee
Similarly, if $\bra{y}\in\mathcal{V}^*$,
$\bra{\tilde{y}}\in\tilde{\mathcal{V}}^*$ then
\be
\bra{y}\otimes\bra{\tilde{y}} \in \mathcal{V}^* \times \tilde{\mathcal{V}}^*
\text{ ,}
\qquad 
(\bra{y}\otimes\bra{\tilde{y}}) (\ket{x}\otimes\ket{\tilde{x}})
= \ip{y}{x} \ip{\tilde{y}}{\tilde{x}} \text{ .}
\ee
Tensor multiplication is associative and distributive over addition, and
$\ket{}^{\otimes n}$ denotes $n$ identical vectors tensor multiplied by each
other.\\

To see how this notation works when applied to stochastic models, consider a
Poisson process. For simple Markov chains like this, it is common to write the
transition rates explicitly, after stating that the system is characterised by
an integer-valued stochastic variable $N$:
\be
N \rightarrow N+1 \text{ at rate } \lambda \text{ .}
\label{explicit}
\ee
The limitation of expressions of the form~\eqref{explicit} is that for more
complex processes, a very long list of events and rates can be generated,
making further manipulations difficult. The Kolmogorov equations for
the Markov chain are an alternative definition:
\be
\frac{d \mathbf{p}}{dt} = \mathbf{p}\mathbf{Q} \text{ , where}\quad 
\mathbf{p} = \left( \begin{array}{c}
\text{Pr}(N=0) \\\text{Pr}(N=1) \\ \text{Pr}(N=2) \\
\vdots \end{array}
\right) \text{ ,} \quad
\mathbf{Q} = \lambda \left( \begin{array}{cccc} 
-1 & 1 & 0 & \cdots\\
0 & -1 & 1 & \cdots\\
0 & 0 & -1 & \cdots\\
\vdots & \vdots & \vdots &  \end{array} \right) \text{ .}
\ee
Again, for complex stochastic processes this can quickly yield objects that are
almost impossible to manipulate. To use Dirac notation, we first define a
state-space and operator
\be
\mathcal{S} = \text{span}%
\left\{ \ket{N} \right\}_{N=0}^{\infty} \text{ ,} \quad
\hat{q} \ket{N} = \ket{N+1}\text{ .}
\ee
Having made these definitions, definition of the Markov chain can then be done
extremely parsimoniously:
\be
\frac{d}{dp} \ket{p} = \hat{Q}\ket{p} \text{ , where} \quad
\hat{Q} = \lambda \left( \hat{q} - \unitop \right) \text{ .}
\label{KFeq}
\ee
Clearly, the extra effort to obtain such an expression is superfluous for a
model as simple as the Poisson process; however for the complex systems we
consider later, alternatives are simply too unwieldy. Since probabilities must
sum to one, we write
\be
\ip{1}{p} = 1\text{ ,} \quad \text{where} \quad
\ket{1} = \sum_{N=0}^{\infty} \ket{N} \text{ ,} \quad \text{and} \quad
\ip{M}{N} = \delta_{M,N} \text{ .}
\ee
Here and throughout, $\delta$ is the Kronecker delta function.  The normalising
`ket' $\ket{1}$ is henceforth used to stand for an unweighted sum over basis
states for the system under consideration.

\subsubsection{Within-subnet dynamics}

Here we consider stochastic epidemics dynamics for a network of size $n$ with
adjacency matrix $\mathbf{a}$.  Our starting point is a node-level state space
\be
 \mathcal{S} = \text{span}%
\left\{ \ket{S}, \ket{I}, \ket{R} \right\} \text{ ,}
\ee
Defined such that, where we use letters $A, B, \ldots$ to represent generic
disease states
\be
 \ip{A}{B} = \delta_{A,B} \text{ .}
\ee
We then define five abstract operators: three that
return the appropriate infection state
\begin{align}
\hat{S} \ket{S} & = \ket{S} , & 
 \hat{S} \ket{I} & = 0 , & 
 \hat{S} \ket{R} & = 0 ,\nonumber \\
\hat{I} \ket{S} & = 0 , & 
 \hat{I} \ket{I} & = \ket{I} , & 
 \hat{I} \ket{R} & = 0 ,\nonumber \\
\hat{R} \ket{S} & = 0, & 
 \hat{R} \ket{I} & = 0 , & 
 \hat{R} \ket{R} & = \ket{R} ;
\end{align}
and two that correspond to transmission and recovery
\begin{align}
\hat{t} \ket{S} & = \ket{I} , & 
 \hat{t} \ket{I} & = 0 , & 
 \hat{t} \ket{R} & = 0 ,\nonumber \\
\hat{r} \ket{S} & = 0 ,& 
 \hat{r} \ket{I} & = \ket{R}  ,& 
 \hat{r} \ket{R} & = 0 .
\end{align}
So a general state under consideration obeys
\be
\ket{p} \in \mathcal{S}^{n} \text{ ,} \quad
\ip{1}{p} = 1 \text{ , where }
\ket{1} := \left( \ket{S} + \ket{I} + \ket{R} \right)^{\otimes n}
\text{ .}
\ee
Where $\hat{\mathcal{O}}$ is an operator defined to act on elements of
$\mathcal{S}$, we define an operator acting on the complete state space using
subscripting so that
\be
\hat{\mathcal{O}}_i := \unitop \otimes \cdots  
 \underbrace{\otimes \hat{\mathcal{O}} \otimes}_{i\text{th place}}
 \cdots \unitop \text{ .}
\ee
Having set up this machinery, we can now write the system's dynamics in an
extremely compact form:
\be
\frac{d}{dt} \ket{p} = \hat{Q} \ket{p} \text{ , where }
 \hat{Q} = \tau \sum_i (\hat{t}_i - \hat{S}_i) \sum_j a_{ij} \hat{I}_j
  + \gamma \sum_i (\hat{r}_i - \hat{I}_i) \text{ .}
 \label{fulldyn}
\ee
Despite this compact expression, the actual dimensionality of the system above
grows extremely quickly with network size for numerical and analytical work.
Methods available for increasing the tractability of these equations,
particularly for final outcomes, are discussed next.

\subsubsection*{Calculation of final outcomes}

We are interested in calculation of the probability that a node $j$ experiences
infection, given an initial probability vector 
\be
\ket{\boldsymbol{\pi}} := \sum_{\mu_1=0}^{1} \cdots \sum_{\mu_{n}=0}^{1}
\prod_{j=1}^{n} \left((\boldsymbol{\pi})_j \hat{S}_j \right)^{1-\mu_j}
\left((1-(\boldsymbol{\pi})_j) \hat{I}_j \right)^{\mu_j}\ket{1} \text{ .}
\label{piinit}
\ee
For the outcomes relevant to working out invasion thresholds like~\eqref{Kl}
the special case of exactly one initial infection is simply a special case:
\be
\ket{o} := \hat{I}_o \prod_{j\neq o} \hat{S}_j \ket{1} \text{ .}
\label{oket}
\ee
An insight from Bailey~\cite{Bailey:1975} is that the probability that a node
$j$ recovers increases at a rate $\gamma$ multiplied by the probability that it
is currently infectious. This means that if~\eqref{fulldyn} holds, for initial
condition~\eqref{piinit}, then
\be
P(j | o) = \int_0^\infty \gamma \bra{I_j} e^{\hat{Q} t} \ket{\boldsymbol{\pi}}
dt \text{ , where} \quad \ket{{I}_j} := \hat{I}_j \ket{1} \text{ ,}
\label{pathint}
\ee
where exponentiation of an operator is defined through the power series
\be
e^{\hat{\mathcal{O}}} = \unitop + \sum_{k=1}^\infty
\frac{\hat{\mathcal{O}}^k}{k!} \text{ .}
\ee
In order to evaluate~\eqref{pathint} efficiently, we can make use of the
general theory of path integrals for Markov chains~\cite{Pollett:2002}.  Since
the presence of absorbing states in~\eqref{pathint} would cause the integral to
be poorly defined, we need first to decompose the state space into an absorbing
set $\mathcal{A}$ and a non-absorbing set $\mathcal{C}$:
\be
\mathcal{S}^{n} = \mathcal{A} \cup \mathcal{C} \text{ ,}
\ee
which can be done through the definition of projection operators
\be
 \hat{P}_{\mathcal{A}}  = \sum_{\{A_i\}_{i=1}^{n} \in \{ S,R \}^{\otimes n}}
 \ket{A_1} \otimes \cdots \otimes \ket{A_n}
 \bra{A_1} \otimes \cdots \otimes \bra{A_n} \text{ ,} \quad
 \hat{P}_{\mathcal{C}}  = \unitop - \hat{P}_{\mathcal{A}} \text{ .}
\ee
The quantities of interest can then be written
\be
P(j|\boldsymbol{\pi}) = \ip{\boldsymbol{\pi}}{z}\text{ ,}
\ee
where 
\be
\bigg( \tau \sum_i (\hat{t}_i^\dagger - \hat{S}_i) \sum_j a_{ij} \hat{I}_j
  + \gamma \sum_i (\hat{r}_i^\dagger - \hat{I}_i)
\bigg) \hat{P}_{\mathcal{C}}
\ket{z} = -\gamma\ket{I_j} \text{ ,} \label{linQeq}
\ee
an equation that uses new operators
\begin{align}
\hat{t}^\dagger \ket{S} & = 0 , & 
 \hat{t}^\dagger \ket{I} & = \ket{S} , & 
 \hat{t}^\dagger \ket{R} & = 0 ,\nonumber \\
\hat{r}^\dagger \ket{S} & = 0 ,& 
 \hat{r}^\dagger \ket{I} & = 0 ,& 
 \hat{r}^\dagger \ket{R} & = \ket{I} .
\end{align}
The final outcome probabilities can therefore be calculated through the solution
of the set of linear equations~\eqref{linQeq}.\\

An alternative is to use the `multi-type' formula from Ball~\cite{Ball:1986}.
This approach has the disadvantage that a set of linear equations must be
solved for each possible initial condition of the form~\eqref{oket}, but the
advantage of being able to include arbitrary distributions of recovery times.
Supposing the probability density function for recovery time is $f(t)$, let
\be
\Psi(s) := \frac{1}{\int_0^\infty e^{-st} f(t) dt} \text{ ,}
\ee
which obeys $\Psi(0) = 1$ in general.  Then for initial probability vector
$\ket{o}$, and final probability vector $\ket{p^\infty_o} \in\mathcal{A}$, the
following equation holds for any $\ket{B_1} \otimes \cdots \otimes \ket{B_n}
\in \mathcal{A}$\,:
\be
\sum_{A_1 \in \{S,B_1\}} \cdots
\sum_{A_o = R} \cdots
\sum_{A_n \in \{S,B_n\}}
\bra{A_1} \otimes \cdots \otimes \ip{A_n}{p^\infty_o} \prod_i \Psi\bigg(\tau
\ip{R_i}{A_i} \sum_j a_{ij} \ip{S_j}{B_j} \bigg) = 1 \text{ .}
\label{ballfs}
\ee
For Markovian recovery at rate $\gamma$, we have $\Psi(s) = (s+\gamma)/\gamma$,
which leads to terms that look much like those obtained during the solution
of~\eqref{linQeq}, and indeed Bailey~\cite{Bailey:1975} notes that equations of
the form~\eqref{ballfs} can be derived from equations of the
form~\eqref{linQeq} in the case of a complete graph.

\subsubsection*{Automorphism-driven lumping}

Recently, the technique of automorphism-driven lumping has been applied to
epidemic dynamics on networks~\cite{Simon:2011} and
percolation~\cite{Karrer:2010}. This approach reduces the complexity of network
problems by making systematic use of discrete symmetries of the network. In
particular, the automorphism group of a graph $G$ of size $n$ with adjacency
matrix $\mathbf{a}$ is a subset of the permutation group: $\mathrm{Aut}(G)
\subseteq \mathrm{S}_n$.  The elements of the automorphism group leave the
adjacency matrix invariant:
\be
 \mathbf{M} \in \mathrm{Aut}(G) \quad \Leftrightarrow \quad
 \mathbf{a} = \mathbf{M} \mathbf{a} \mathbf{M}^{\top} \text{ .}
\ee
The use of this insight to lump epidemic equations requires some
care in the labelling of dynamical variables~\cite{Simon:2011}. Using the
notation above, we relabel a generic dynamical state of the system
\be
 \ket{A_1} \otimes \cdots \otimes \ket{A_n} \equiv 
 \ket{ \{ (A_1,1) , \ldots , (A_n,n) \} } \text{ ,}
\ee
i.e.\ we go from an ordered set of states to an unordered set of pairs of
states and node numbers. `Lumped' basis states for the dynamical
system~\eqref{fulldyn} can then be defined according to the orbits of the
automorphism group---this means that states like the above are lumped
together into classes like
\be
 L(A_1,\ldots,A_n) =
 \{ \ \{ (A_1,M(1)) , \ldots , (A_n,M(n)) \} \ | \
 \mathbf{M} \in \mathrm{Aut}(G) \ \}
 \text{ ,}
\ee
where $M(i)$ is the index of the non-zero component of the $i$-th row of the
permutation matrix $\mathbf{M}$. The dynamical equivalence of these states can
be seen by repeated substitution of $\mathbf{a} \rightarrow \mathbf{M}
\mathbf{a} \mathbf{M}^{\top}$ into~\eqref{fulldyn}. Clearly, lumping
classes must contain states that all have the same eigenvalues of $\hat{S}$ and
$\hat{I}$; and in the limiting case of a fully connected graph such that
$\mathrm{Aut}(G) = \mathrm{S}_n$, only these aggregate eigenvalues are required
to describe the system~\cite{Simon:2011}.

\subsubsection{Global dynamics}

Recently, a set of dynamics was presented that are a manifestly asymptotically
exact description of the mean behaviour of an \SIR{} epidemic on a
configuration-model network~\cite{Ball:2008} (equivalent to a stubby subnet
model where all subnets have one node). The idea behind this construction is
that, once numbers of half-links are allocated in the configuration model, then
the formation of full links is a Poisson process that can be compounded with
the epidemic process. Individuals therefore start the epidemic with half links,
and the network is constructed at the same time as the epidemic tree, reducing
the number of half-links in the system over time.

We now write this construction in Dirac vector space notation, so that this
approach may be readily combined with the within-subnet dynamics above to
define global dynamics.  Our starting point is a set of states that represent a
number of `remaining half-links'
\be
\mathcal{S} = \text{span}%
\left\{ \ket{l} \right\}_{l = 0}^{k_{\mathrm{max}}}
 \text{ , such that } \ip{l'}{l} = \delta_{l,l'} \text{ ,}
\ee
where $k_{\mathrm{max}}$ is the maximum node degree (or more generally maximum
number of stubs). We define two operators on such states: a link number
operator, and a link-number lowering operator:
\be
\hat{l} \ket{l} = l \ket{l} \text{ ,} \qquad
\hat{l}^{-} \ket{l} = \begin{cases}
\ket{l-1} & \text{ if } l \geq 1 \text{ ,}\\
0 & \text{ otherwise.}
\end{cases}
\ee
We now consider how remaining half-links interact with disease state. These are
taken as a tensor product,
\be
 \ket{A,l} = \ket{A} \otimes \ket{l} \text{ , so that }
 \ip{B,l'}{A,l} = \delta_{A,B} \delta_{l,l'} \text{ .}
\ee
By construction, however, recovered individuals lose all their half-links,
so the state space for this system is
\be
\mathcal{S} = \text{span}%
\left\{ \ket{S,l}, \ket{I,l}, \ket{R,0} 
  \right\}_{l=0}^{k_{\mathrm{max}}} \text{ .}
\ee
We then define four operators on this space, which we present in terms of
their non-trivial action
\ba
\hat{t} \ket{S,l} & := \left( \hat{t} \ket{S} \right) \otimes \ket{l}
  = \ket{I,l} \text{ ,}\\
\hat{b} \ket{S,l} & := \left( \hat{t} \ket{S} \right) \otimes 
 \left( \hat{l}^{-} \ket{l} \right) = \ket{I,l-1} \text{ ,}\\
\hat{l}^{-} \ket{A,l} & := \ket{A} \otimes 
 \left( \hat{l}^{-} \ket{l} \right) = \ket{A,l-1} \text{ ,}\\
\hat{r} \ket{I,l} & := \ket{R,0} \text{ .}
\ea
Three of these operators are simple uplifts, but the operator $\hat{b}$ for
global infection is new.  To define the dynamics of this system, we start with
a general state
\be
\ket{p} = \sum_{l=0}^{k_{\text{max}}}%
 \left( x_l(t) \ket{S,l} + y_l(t) \ket{I,l} \right)
  + z(t) \ket{R,0} \text{ ,}
\ee
which obeys
\be
\ip{1}{p} = 1 \text{ , for } \ket{1} := \sum_{l=0}^{k_{\text{max}}}%
 \left( \ket{S,l} + \ket{I,l} \right) + \ket{R,0} \text{ .}
\ee
There is also a non-linear term for the density of infection amongst free
half-links that appears in the system,
\be
\rho[p] := \frac{\bra{1} \hat{I} \hat{l} \ket{p}}{\bra{1}
 \hat{l} \ket{p}} \text{ .}
\ee
Then a representation of the expected \SIR{} dynamics on a configuration-model
network is given by
\ba
\hat{Q}[p] & := \gamma \left( \hat{r} - \hat{I} \right)
  + \tau \left(\hat{l}^{-} - \unitop\right) \hat{l} \hat{I}
  + \rho[p] \left(\gamma + \tau \right)\left(\hat{l}^{-} - \unitop \right)
   \hat{l}
  + \rho[p] \tau \left(\hat{b} - \hat{S} \right) \hat{l} \text{ ,}\\
\frac{d}{dt} \ket{p} & = \hat{Q}[p] \ket{p} \text{ .}
\ea
The significance of these dynamics is that they do not grow in dimension with
network size; in fact, they are asymptotically exact in the large system-size
limit, which is inaccessible through simulation or direct integration
of~\eqref{fulldyn}.

\subsubsection{Full system dynamics}

For a network made up of stubby subnets, it is possible to a make the same
construction as above, where global links are made along with the epidemic
process. In this case, a general state can be written
\be
\ket{p} = \sum_{\sigma, A_1, \ldots, A_{n_\sigma}, l_1,\ldots, l_{n_\sigma}}
 {{p_{\sigma}}^{A_1 \ldots A_{n_\sigma}}}_{l_1 \ldots l_{n_\sigma}}(t)
 \ket{\sigma}\otimes\ket{A_1,l_1} \otimes \cdots \otimes 
 \ket{A_{n_\sigma},l_{n_\sigma}} 
 \text{ ,}
\ee
where $\ip{\bar{\sigma}}{\sigma} = \delta_{\bar{\sigma},\sigma}$ as would
be expected.  Clearly, any attempt to write down differential equations for the
tensor representation of this system, ${{p_{\sigma}}^{A_1 \ldots A_n}}_{l_1
\ldots l_n}(t)$, will involve extremely complex expressions. By contrast, using
the formalism of Dirac notation and operators that we have developed above, we
can write the dynamics for this system as
\ba
\hat{P}_\sigma & := \sum_{A_1, \ldots, A_{n_\sigma}, l_1,\ldots, l_{n_\sigma}}
 \ket{\sigma}\otimes\ket{A_1,l_1} \otimes \cdots \otimes 
 \ket{A_{n_\sigma},l_{n_\sigma}} \bra{A_{n_\sigma},l_{n_\sigma}} \otimes \cdots
	\otimes\bra{A_1,l_1} \otimes \bra{\sigma} \\
\rho[p] & := \frac{\bra{1} 
 \sum_\sigma \sum_{i=1}^{n_\sigma} \hat{I}_i \hat{l}_i \hat{P}_\sigma
 \ket{p}}{\bra{1}
  \sum_\sigma \sum_{i=1}^{n_\sigma} \hat{l}_i \hat{P}_\sigma \ket{p}} 
 \text{ ,} \\
\hat{Q}_{\sigma}[p] & := \gamma \sum_{i=1}^{n_\sigma} \left( \hat{r}_i 
  - \hat{I}_i \right)
  + \tau \sum_{i=1}^{n_\sigma} \left(\hat{l}^{-}_{i} - \unitop\right) \hat{l}_i
 \hat{I}_i \\ & \quad
  + \tau \sum_{i=1}^{n_\sigma} \left(\hat{t}_i - \hat{S}_i\right) 
  \sum_{j=1}^{n_{\sigma}} a^{\sigma}_{ij} \hat{I}_j
  \\ & \quad
  + \rho[p] \left(\gamma + \tau \right) \sum_{i=1}^{n_\sigma}
  \left(\hat{l}^{-}_{i} - \unitop \right) \hat{l}_i
  + \rho[p] \tau \sum_{i=1}^{n_\sigma} \left(\hat{b}_i - \hat{S}_i \right) 
 \hat{l}_i \text{ ,}\\
\frac{d}{dt} \ket{p} & = \sum_{\sigma}
 \hat{Q}_{\sigma}[p] \hat{P}_{\sigma} \ket{p} \text{ .}
\label{exeqs}
\ea
These equations have the same significance as above: the asymptotically exact
expected epidemic dynamics of a class of clustered dynamics can be calculated
for the large system-size limit of a network.

\section{Examples}

We now turn to some examples of the methodology presented above to specific
networks. Throughout this section we work in natural units such that the
recovery rate $\gamma =1$.

\subsection{Invasion and final size}

We consider invasion on the two locales shown in
Figure~\ref{fig:envelope}(a,b).  These networks are constructed from the
envelope / diamond motif as shown, so that every individual has exactly $n$
links. This means that all differences between this model and an $n$-regular
random graph derive from the presence and structure of short loops in the
network and not heterogeneity in node degree.  The locale basic reproductive
ratio is given (after straightforward but tedious manipulations probably best
carried out using a computer algebra system) by:
\begin{multline}
R_L =
\big(\tau  \big(2 (n-3)^2 +(n (25 n-142)+204) \tau +(n (133
n-716)+982) \tau^2\\ 
+(n (377 n-1948)+2570) \tau^3+(n (563
n-2846)+3672) \tau^4\\
+2 (n (193 n-968)+1239) \tau^5+12 (8
(n-5) n+51) \tau^6\big)\big)\\
/\big((2 n-5) (1 +\tau )^4 (1 +2 \tau )^2
(1 +3 \tau )\big) \text{ .}
\end{multline}
It is also possible to compare this envelope-based network to other networks
that are also $n$-regular, but have different generalised clustering. This
comparison is shown for final sizes where $n=4$ in Plot (c) of
Figure~\ref{fig:envelope}; as would be expected, networks with more short
loops are harder for a disease to invade.

The question might also be asked as to how quickly epidemics simulated using
Monte Carlo methods on finite networks converge to the asymptotic results,
which is considered in Figure~\ref{fig:envelope}(d,e,f).  These show that even
for networks of a few thousand nodes, asymptotic results provide a useful guide
to expected behaviour.

\subsection{Full Dynamics}

\begin{figure}
\centering
\subfloat[]{
\scalebox{0.275}{\resizebox{\textwidth}{!}{%
\includegraphics{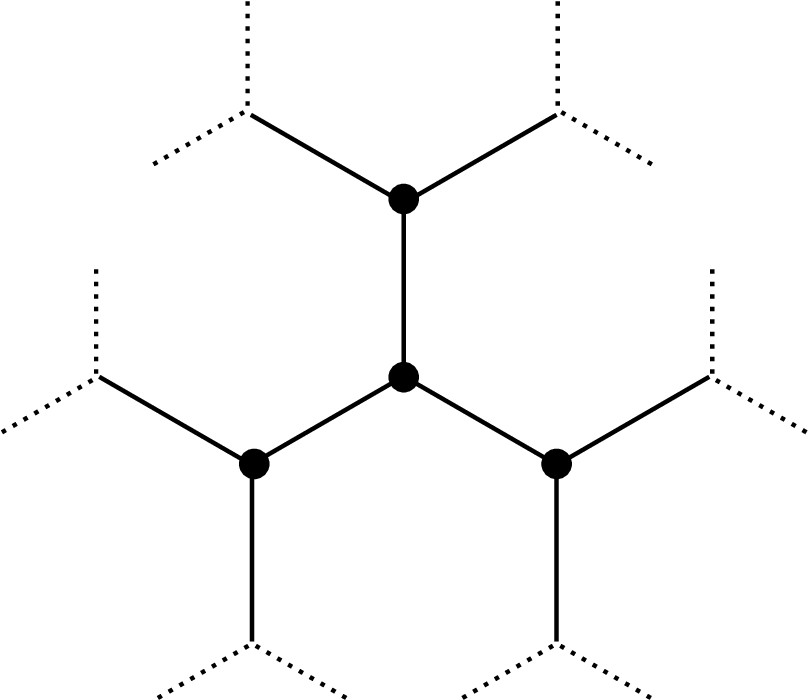}}}\hspace{0.075\textwidth}} \quad
\subfloat[]{
\scalebox{0.35}{\resizebox{\textwidth}{!}{%
\includegraphics{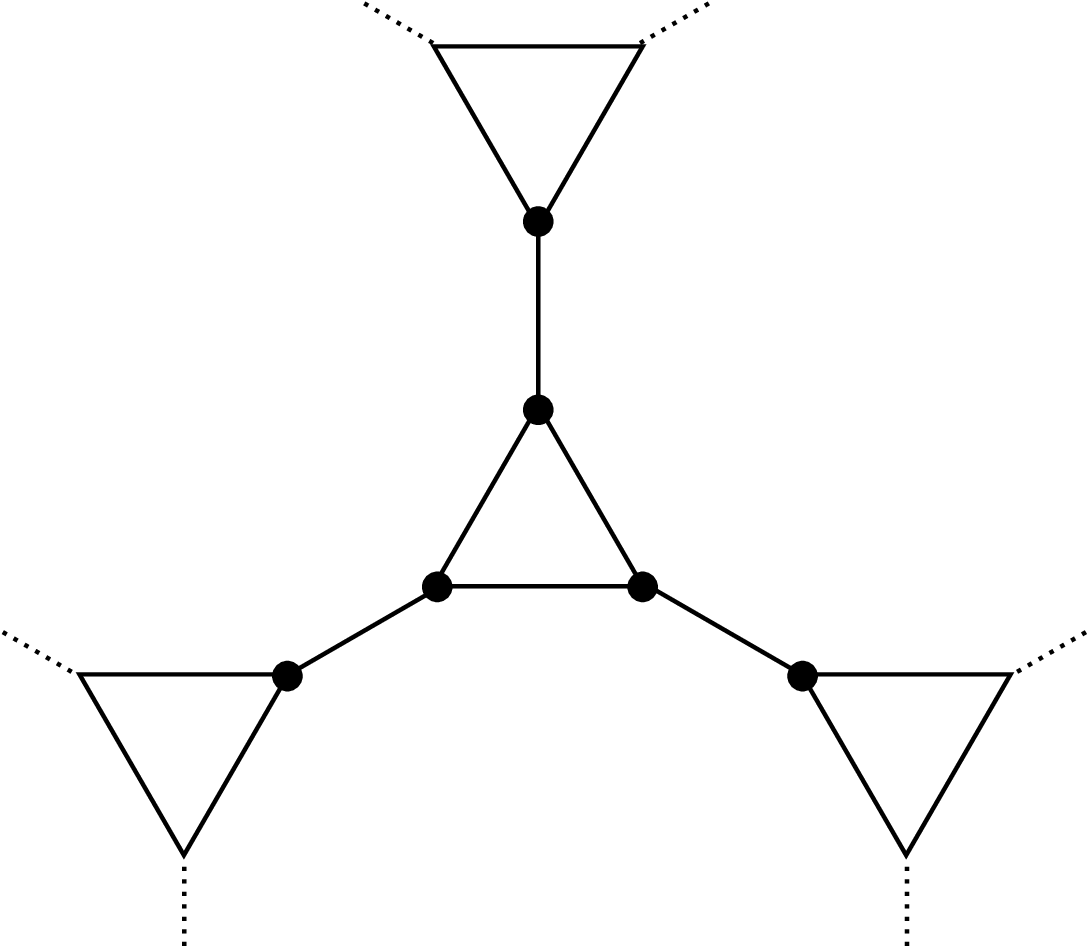}}}} \\ 
\subfloat[]{
\scalebox{0.35}{\resizebox{\textwidth}{!}{%
\includegraphics{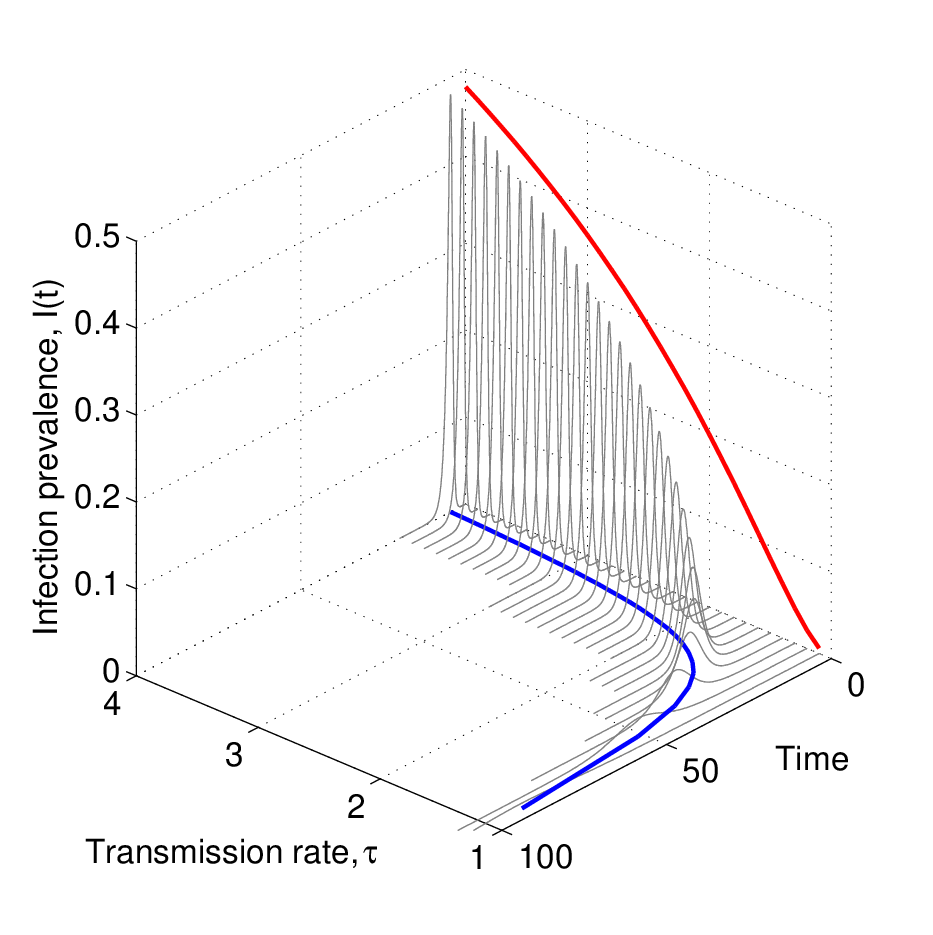}}}} \quad
\subfloat[]{
\scalebox{0.35}{\resizebox{\textwidth}{!}{%
\includegraphics{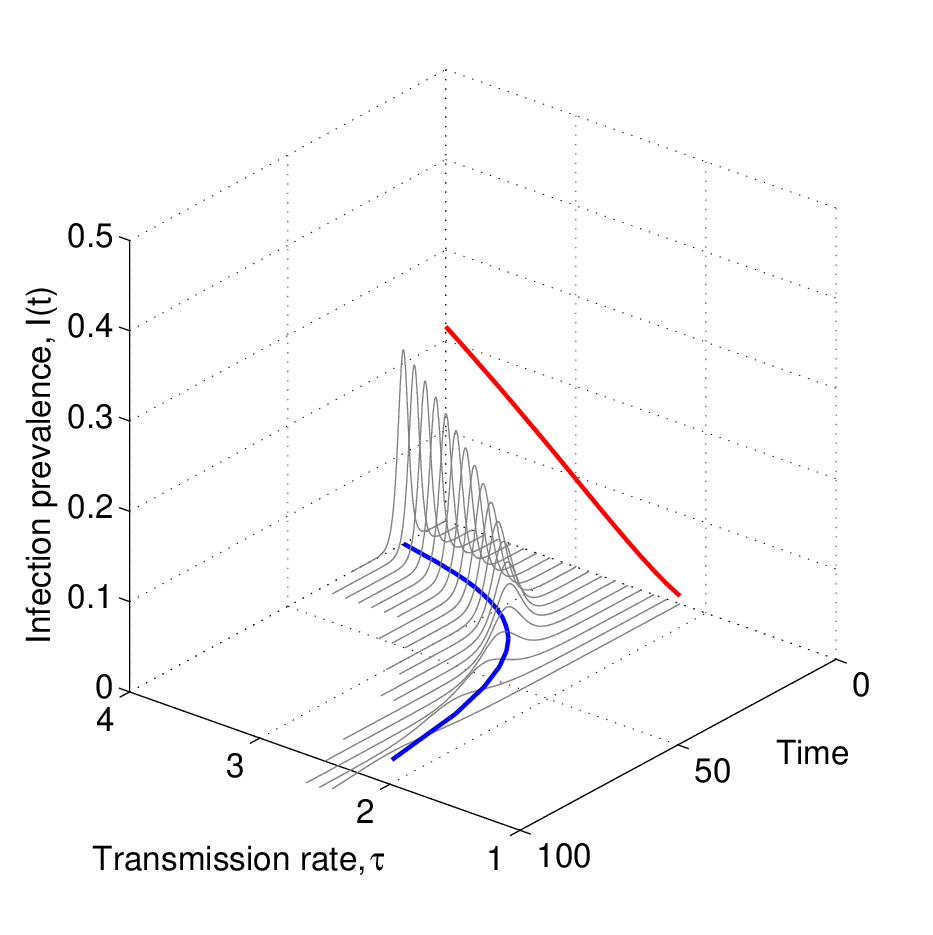}}}}\\
\subfloat[]{
\scalebox{0.35}{\resizebox{\textwidth}{!}{%
\includegraphics{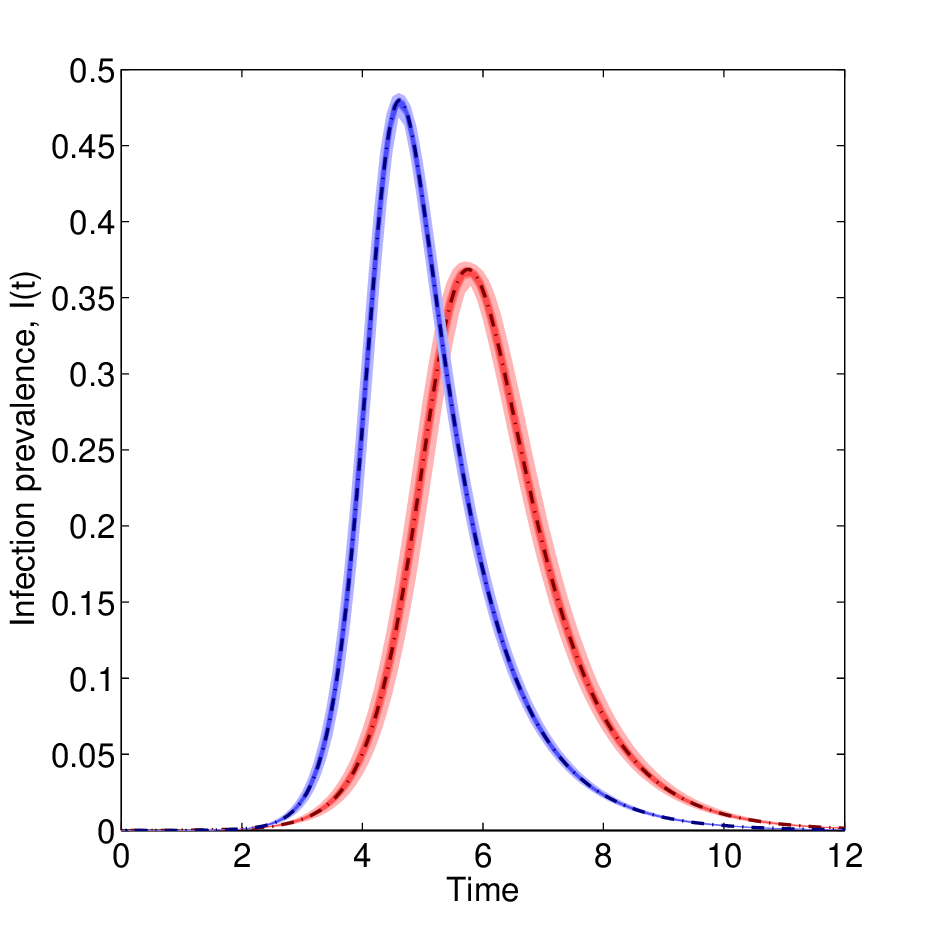}}}} \quad
\subfloat[]{
\scalebox{0.35}{\resizebox{\textwidth}{!}{%
\includegraphics{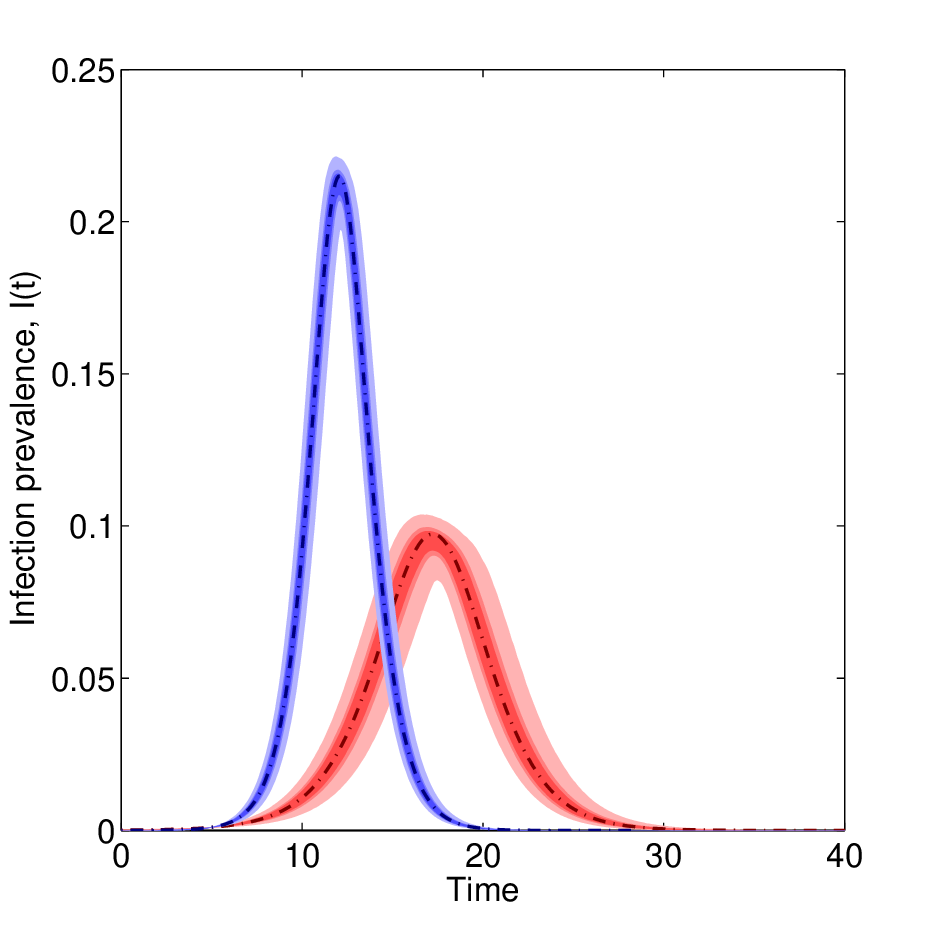}}}}
\caption{Exact transient epidemic dynamics for two special networks. (a) shows
a typical location in the unclustered graph, and (b) shows a typical location
in the clustered graph. Epidemic curves (grey) for different parameter values
are shown in (c), (d) respectively. Peak times (blue) and peak heights (red)
are projected onto the appropriate axes. (e) and (f) compare the asymptotic
results (dot-dashed lines) to simulations on networks of size $N=10^4$ for
$\tau=3$ (red) and $\tau=4$ (blue). Simulation prediction intervals at 50\%,
67\% and 95\% are shown as shaded regions.}
\label{fig:transient}
\end{figure}

While invasion thresholds are of practical interest, transient dynamical
features of epidemics are also important, and are not always simply determined
by consideration of thresholds. Figure~\ref{fig:transient} shows the transient
behaviour in the large system-size limit for two special graphs, both of which
3-regular: (a) a configuration-model network where each node has 3 stubs; (b) a
stubby-subnet graph composed of triangles with each node having one stub. The
dynamics as defined above give the epidemic curves shown in (c) for the CM
network and (d) for the triangle-based network respectively.  As above, we
consider the relationship between these results and direct stochastic
simulation, with results shown in (e) and (f). Clearly, convergence is fast for
epidemics with a transmission rate much greater than the threshold, but near
threshold stochastic, finite-size effects are much more significant.

\section{Other solvable networks}

It has been clear for some time that a network (or otherwise structured
population) with a local-global distinction will admit a solution to an \SIR{}
epidemic on that network~\cite{Ball:2002}. As a practical adjunct to this, both
the local and global features of the network must individually admit solution.
The stubby-subnet networks here propose one such distinction: each node can be
uniquely assigned to a local unit of clustered structure; and global mixing
happens through a configuration model network.

We now consider three other versions of this concept, firstly by introducing
assortative mixing outside the subnet, secondly using the recently defined
role-based networks, and finally to weighted networks. 

\subsection{Assortativity}

In~\cite{Newman:2002}, a generalisation of the configuration model was
developed to incorporate the notion of assortativity. Such assortativity (or
even disassortativity) is a mainstay of epidemiology, and much theoretical
effort has been expended to model its effects~\cite{Diek:2000}. To describe
assortativity, we introduce a correlation matrix ${C}_{\bar{\lambda},\lambda}$
(analogous to the $e_{kl}$ of~\cite{Newman:2002}) that multiplies the
probabilities that two locales are linked globally compared to the
configuration model. For such a network, the locale next generation matrix is
\be \mathcal{K}^{L}_{\bar{\lambda} \lambda} = 
T \frac{M_{\bar{\sigma}} s^{\bar{\sigma}}_{\bar{o}}}{%
 s_{\mathrm{tot}}}
 \left( \left(s_o -1\right)C_{\bar{\lambda}, \lambda}
  + \sum_{j \in v_\sigma \ominus o} P_\sigma (j | o) s_j 
 C_{\bar{\lambda}, ( \sigma, j )} \right)
 \text{ ,}
\ee
and an appropriate threshold parameter will be given by the dominant eigenvalue
of this matrix. Exact transient dynamics for such a system should also be
straightforward to write down: in addition to indexing a node with its
effective remaining half-links and disease state, each node should also be
indexed by locale.  Instead of having homogeneous transmission on the basis of
pairing half-links at rate $\tau$, the rate should then be multiplied by
${C}_{\bar{\lambda},\lambda}$. Of course, this yields equations that are at
least quadratic rather than linear in maximum node degree, making numerical
integration correspondingly more difficult.

\subsection{Role-based networks}

Role-based networks as considered in~\cite{Newman:2009,Miller:2009,Karrer:2010}
involve a different definition of local and global. In these networks, it is
links that can be uniquely assigned to a local unit of clustered structure,
meaning that nodes can be attached to many different clustered subgraphs.  This
clearly allows a next-generation matrix to be established by indexing cases by
the unit of structure through which they acquired infection, as
in~\cite{Miller:2009}.  The definition of asymptotically exact dynamics is less
clear in this case, however dynamical approaches such as~\cite{Volz:2011} that
are in extremely good numerical agreement with simulation, and may turn out to
be large system-size limits through further work, can clearly be extended to
role-based networks.  The primary differences between stubby-subnet and
role-based networks are that the former can specify an exact structure of stubs
for each node in a clustered motif, while the latter can involve each node in
several motifs. As such, these are best seen as complementary approaches to the
fast-moving field of solvable clustered networks.

\subsection{Weighted networks}

While all networks discussed above have been topological (i.e.\ links are
either present or not) all of the analysis above carries through exactly if
within-subnet links are weighted, so that values $a^{\sigma}_{ij} \in
\mathbb{R}_+$ can be substituted into e.g.~\eqref{exeqs}. It is also possible to
stratify global links into multiple contexts, each with a given strength (i.e.\
different values of $T$) although this latter modification does increase the
system dimensionality, while weighting within-subnet dynamics does this only if
the weighting breaks a discrete symmetry of the topological network.

\section{Discussion}

This paper has presented a manifestly asymptotically exact way to calculate
transient epidemic dynamics on a class of clustered networks. As such, it
complements existing work based on simpler network structure or moment closure;
however, this is done at the cost of extremely high dimensional ODE systems,
with even the simplest triangle-based example above involving 216 equations. Of
course, computational resources continue to improve, and so the possibility of
considering more complex networks and disease natural history cannot be ruled
out, but is not currently easily done. At present, there is no perfect
technique for the consideration of epidemic dynamics on networks: direct
simulation is versatile, but hard to interpret; moment closure is numerically
fast but the criteria under which it is accurate are currently unclear; and
analytic approaches are either extremely high dimensional or restricted to
special network types.  In particular, local tree-like structure of some form
has been a feature of all analytic approaches to date, and may be an
indispensable assumption.  This means that there is merit to development of all
available approaches, and it is hoped that this paper broadens the range of
networks on which certain epidemiological results can be computed, contributing
to our understanding of the fast-moving but complex field of contact network
epidemiology.

\section*{Acknowledgements}

Work funded by the UK Engineering and Physical Sciences Research Council. The
author would like to thank Matt Keeling and Josh Ross for helpful discussions
and comments on this work.


\newpage

\begin{thebibliography}{10}

\bibitem{Bailey:1975}
N.~Bailey.
\newblock {\em The Mathematical Theory of Infectious Diseases and its
  Applications}.
\newblock Charles Griffin and Company, London, 1975.

\bibitem{Ball:1986}
F.~Ball.
\newblock A unified approach to the distribution of total size and total area
  under the trajectory of infectives in epidemic models.
\newblock {\em Advances in Applied Probability}, 18(2):289--310, 1986.

\bibitem{Ball:2002}
F.~Ball and P.~Neal.
\newblock A general model for stochastic {{SIR}} epidemics with two levels of
  mixing.
\newblock {\em Mathematical Biosciences}, 180:73--102, Jan 2002.

\bibitem{Ball:2008}
F.~Ball and P.~Neal.
\newblock Network epidemic models with two levels of mixing.
\newblock {\em Mathematical Biosciences}, 212(1):69--87, Jan 2008.

\bibitem{Ball:2009}
F.~Ball, D.~Sirl, and P.~Trapman.
\newblock Threshold behaviour and final outcome of an epidemic on a random
  network with household structure.
\newblock {\em Advances in Applied Probability}, 41(3):765--796, Jan 2009.

\bibitem{Ball:2010}
F.~Ball, D.~Sirl, and P.~Trapman.
\newblock Analysis of a stochastic {SIR} epidemic on a random network
  incorporating household structure.
\newblock {\em Mathematical Biosciences}, 224(2):53--73, Apr 2010.

\bibitem{Bollobas:2011}
B.~Bollob\'{a}s, S.~Janson, and O.~Riordan.
\newblock Sparse random graphs with clustering.
\newblock {\em Random Structures \& Algorithms}, 38(3):269--323, 2011.

\bibitem{Dennery:1967}
P.~Dennery and A.~Krzymicki.
\newblock {\em Mathematics for Physicists}.
\newblock Harper and Row, 1967.

\bibitem{Diek:2000}
O.~Diekmann and J.~Heesterbeek.
\newblock {\em Mathematical Epidemiology of Infectious Diseases: Model
  Building, Analysis and Interpretation}.
\newblock J Wiley, 2000.

\bibitem{Dodd:2009}
P.~J. Dodd and N.~M. Ferguson.
\newblock A many-body field theory approach to stochastic models in population
  biology.
\newblock {\em PLoS ONE}, 4(9):e6855, 09 2009.

\bibitem{Durrett:2007}
R.~Durrett.
\newblock {\em Random Graph Dynamics}.
\newblock Cambridge University Press, 2007.

\bibitem{Gleeson:2009}
J.~P. Gleeson.
\newblock Bond percolation on a class of clustered random networks.
\newblock {\em Physical Review E}, 80(3):036107, Sep 2009.

\bibitem{Gleeson:2009a}
J.~P. Gleeson and S.~Melnik.
\newblock Analytical results for bond percolation and $k$-core sizes on
  clustered networks.
\newblock {\em Physical Review E}, 80(4):046121, Oct 2009.

\bibitem{House:2009}
T.~House, G.~Davies, L.~Danon, and M.~J. Keeling.
\newblock A motif-based approach to network epidemics.
\newblock {\em Bulletin of Mathematical Biology}, 71:1693--1706, Apr 2009.

\bibitem{Karrer:2010}
B.~Karrer and M.~E.~J. Newman.
\newblock Random graphs containing arbitrary distributions of subgraphs.
\newblock {\em Physical Review E}, 82(6):066118, Dec 2010.

\bibitem{Keeling:1999}
M.~J. Keeling.
\newblock The effects of local spatial structure on epidemiological invasions.
\newblock {\em Proceedings of the Royal Society B}, 266(1421):859--67, Apr
  1999.

\bibitem{Miller:2009}
J.~Miller.
\newblock Percolation and epidemics in random clustered networks.
\newblock {\em Physical Review E}, 80(2):020901, Aug 2009.

\bibitem{MolloyReed}
M.~Molloy and B.~Reed.
\newblock A critical point for random graphs with a given degree sequence.
\newblock {\em Random Structures \& Algorithms}, 6(2/3):161--179, 1995.

\bibitem{Neuts:1975}
M.~F. Neuts.
\newblock Probability distributions of phase type.
\newblock In {\em Liber amicorum {{P}}rofessor emeritus {{Dr. H. Florin}}},
  pages 173--206. Katholieke Universiteit Leuven, Departement Wiskunde, 1975.

\bibitem{Newman:2009}
M.~Newman.
\newblock Random graphs with clustering.
\newblock {\em Physical Review Letters}, 103(5):1--4, Jul 2009.

\bibitem{Newman:2002}
M.~E.~J. Newman.
\newblock Assortative mixing in networks.
\newblock {\em Physical Review Letters}, 89(20):208701, Jan 2002.

\bibitem{Pollett:2002}
P.~K. Pollett and V.~E. Stefanov.
\newblock Path integrals for continuous-time {Markov} chains.
\newblock {\em Journal of Applied Probability}, 39:901--904, 2002.

\bibitem{Ross:2010}
J.~V. Ross, T.~House, and M.~J. Keeling.
\newblock Calculation of disease dynamics in a population of households.
\newblock {\em PLoS ONE}, 5:e9666, Jan 2010.

\bibitem{Simon:2011}
P.~Simon, M.~Taylor, and I.~Kiss.
\newblock Exact epidemic models on graphs using graph-automorphism driven
  lumping.
\newblock {\em Journal of Mathematical Biology}, 62:479--508, 2011.

\bibitem{Trapman:2007}
P.~Trapman.
\newblock On analytical approaches to epidemics on networks.
\newblock {\em Theoretical Population Biology}, 71:160--173, March 2007.

\bibitem{Volz:2011}
E.~M. Volz, J.~C. Miller, A.~Galvani, and L.~Ancel~Meyers.
\newblock Effects of heterogeneous and clustered contact patterns on infectious
  disease dynamics.
\newblock {\em PLoS Computational Biology}, 7(6):e1002042, June 2011.

\end{thebibliography}

\begin{landscape}

\thispagestyle{empty}
\begin{figure}
\centering
\subfloat[]{\hspace{0.025\textwidth}
\scalebox{0.3}{\resizebox{\textwidth}{!}{%
\includegraphics{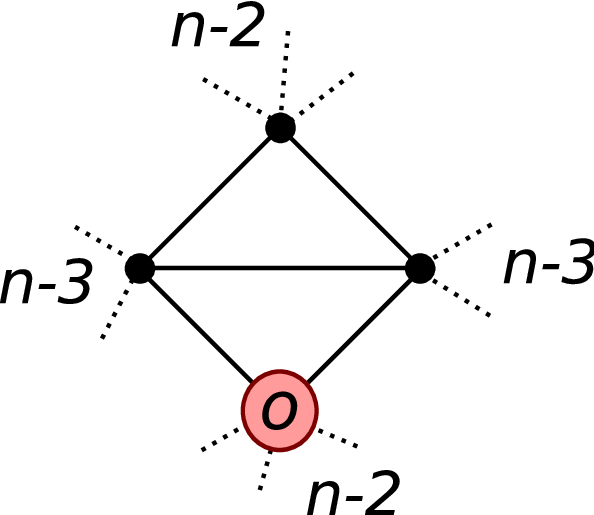}}}
}\quad
\subfloat[]{
\scalebox{0.3}{\hspace{0.1\textwidth}
\resizebox{\textwidth}{!}{%
\includegraphics{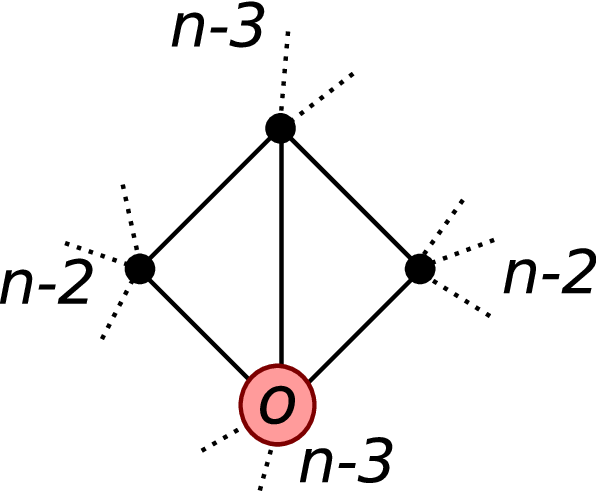}}}
}\quad
\subfloat[]{\scalebox{0.35}{\resizebox{\textwidth}{!}{%
\includegraphics{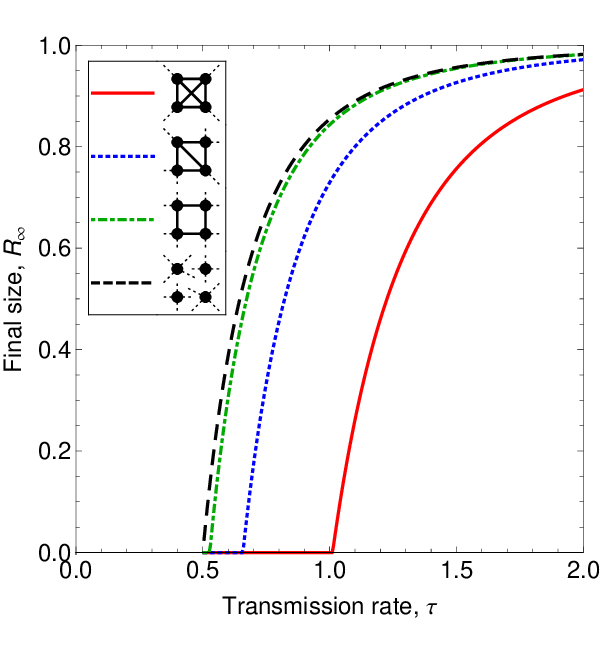}}}} \\
\subfloat[]{\scalebox{0.35}{\resizebox{\textwidth}{!}{%
\includegraphics{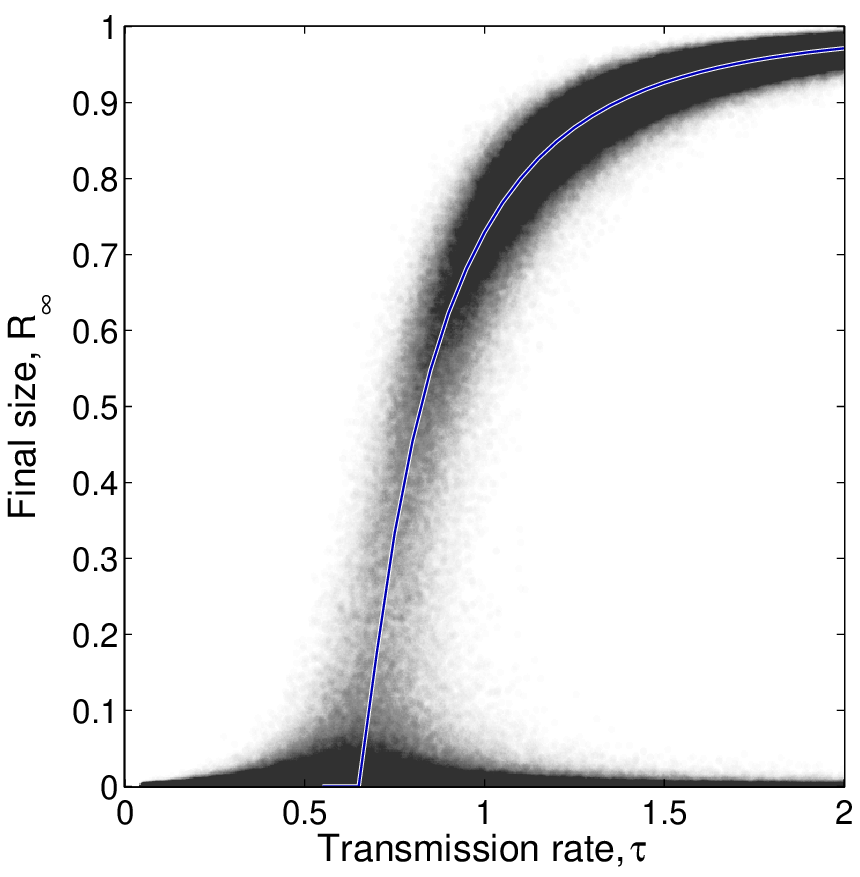}}}} \quad
\subfloat[]{\scalebox{0.35}{\resizebox{\textwidth}{!}{%
\includegraphics{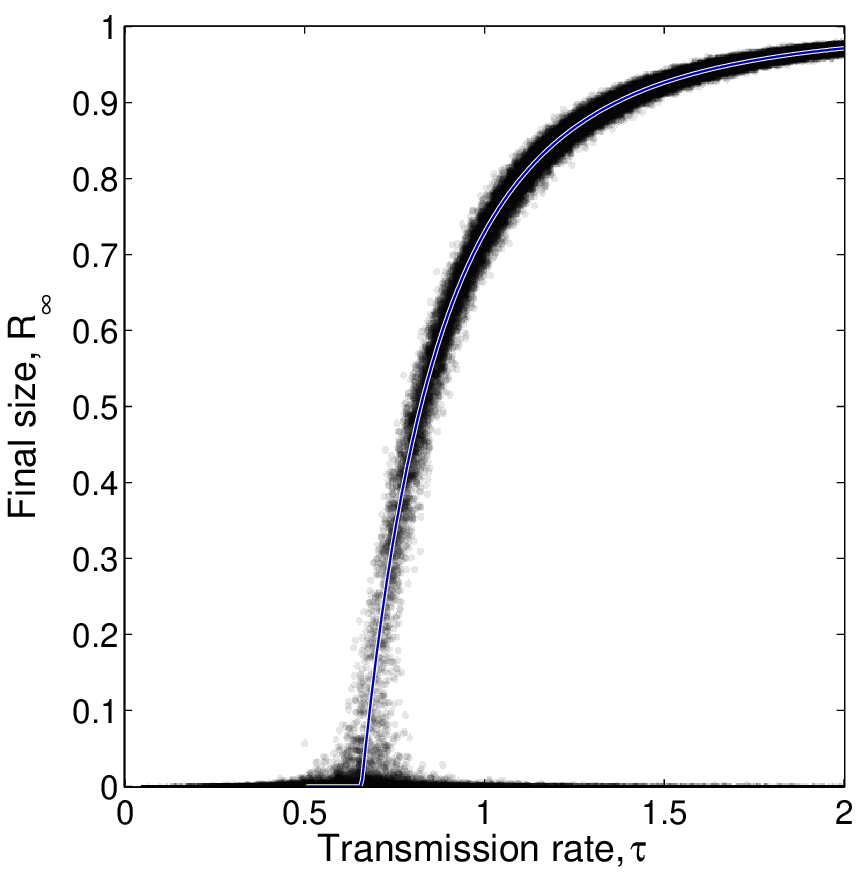}}}} \quad
\subfloat[]{
\scalebox{0.35}{\resizebox{\textwidth}{!}{%
\includegraphics{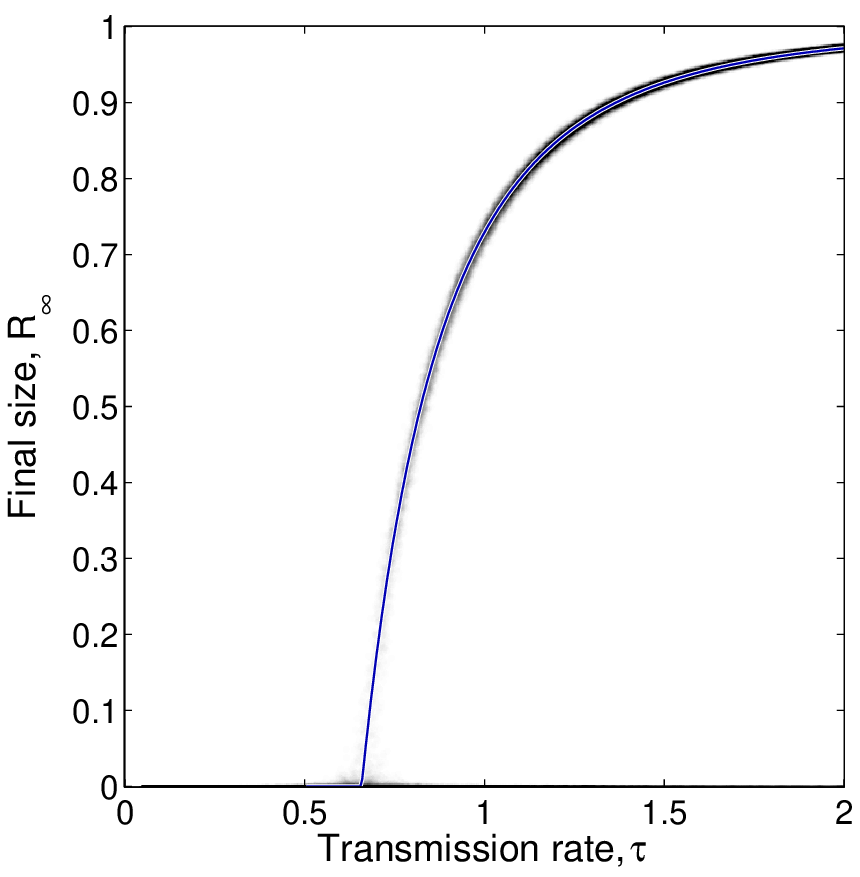}}}}
\caption{Epidemic final sizes on regular graphs with different generalised
clustering. (a) and (b) show the two locales involved in general
`envelope'-based graphs. For $n=4$, $\gamma = 1$, (c) shows asymptotic final
sizes for four different graphs, while remaining plots illustrate the rate of
convergence to the asymptotic result by showing final sizes for $M$ runs on
envelope-based networks of size $N$ where (d) $M=10^6, N=10^3$, (e) $M=10^5,
N=10^4$, (f) $M=10^5, N=5\times{}10^4$.  Each dot represents a realisation, and
blue lines are asymptotic results.}
\label{fig:envelope}
\end{figure}

\end{landscape}

\thispagestyle{empty}

\end{document}